\documentclass[10pt,superscriptaddress,twocolumn,amsmath,amssymb,aps,prl]{revtex4-1}
\usepackage{mathrsfs}
\usepackage{graphicx}
\usepackage{dcolumn}
\usepackage{bm}

\newcommand{\be}{\begin{equation}}
\newcommand{\ee}{\end{equation}}
\newcommand{\bea}{\begin{eqnarray}}
\newcommand{\eea}{\end{eqnarray}}

\begin{document}
\title{Giant Enhancement of the Thermo-Particle Transport Figure of Merit   and Breakdown of the Wiedemann-Franz Law in Unitary Fermi Gases}
\author{Xinloong Han}
\affiliation{Beijing National Laboratory for Condensed Matter Physics, and Institute of Physics, Chinese Academy of Sciences, Beijing 100190,China}
\affiliation{University of Chinese Academy of Sciences, Beijing 100049, China}
\author{Boyang Liu} \email{boyangleo@gmail.com} \affiliation{Institute of Theoretical Physics, Beijing University of Technology
Beijing 100124, China}
\author{Jiangping Hu}\email{jphu@iphy.ac.cn}
\affiliation{Beijing National Laboratory for Condensed Matter Physics, and Institute of Physics, Chinese Academy of Sciences, Beijing 100190,China}
\affiliation{University of Chinese Academy of Sciences, Beijing 100049, China}
\affiliation{Collaborative Innovation Center of Quantum Matter, Beijing, China}

\date{\today}
\begin{abstract}
We theoretically investigate the thermo-particle transport properties of an unitary Fermi gas between two reservoirs connected by a quantum point contact.   We find several distinguished properties that are qualitatively different from those of weak or non-interacting gas systems.  The particle transport figure of merit is drastically  enhanced in  the unitary regime and it  increases as  the transmission coefficient increases,  exactly opposite to the behavior in the weak  or non-interacting gas systems.  The Lorentz number  violates the Wiedemann-Franz law, demonstrating the breakdown of Fermi liquid.
These transport properties are the hallmarks  of  the unitary Fermi gas and are attributed to  the existence of  preformed Cooper pairs. \end{abstract}
 \maketitle

\textit{Introduction.}
Thermo-transport and  thermoelectricity  are critical physical properties of solid state systems\cite{Goldsmid,MacDonald}.  Recently, the study of the transport in cold atom systems has  attracted tremendous research attentions \cite{Stadler,Brantut,Krinner1,Valtolina,Husmann,Krinner2,Hausler,Husmann2018,Lebrat,Burchianti,Sommer,Bardon,Koschorreck,Luciuk,Valtolina2017}. Of particular interests is the realization of thermo-particle transport of cold atoms \cite{Brantut,Husmann2018}. Although atoms are charge-neutral, the conversion between the heat and particle current, as an analogy to thermoelectricity, can be a potential  important application of the ultracold atom systems.

Particularly, in experiment, the transport with two-terminal setup realized by Esslinger's group at ETH is a milestone in quantum simulation \cite{Stadler,Brantut,Krinner1,Husmann,Krinner2,Hausler}. In 2013 Esslinger's group realized the conversion of heat and matter in a non-interacting fermionic cold atom gas and investigated the efficiency of energy conversion by controlling the geometry and disorder strength \cite{Brantut}. The experiment paved a road to study the thermo-particle transport effect in cold atom physics. Furthermore, most recently,  Esslinger's group studied the thermo-particle transport effect in a unitary Fermi gas in the superfluid state and various transport coefficients were measured \cite{Husmann2018}. The experiment suggested that the Wiedemann-Franz law was not valid in the unitary case.  In the unitary regime,  the particle conductivity has been investigated  both experimentally \cite{Stadler,Krinner2} and theoretically \cite{Liu2014,Glazman,Liu2017,Uchino}. However the thermo-particle transport has not been well studied theoretically.

In this Letter, we  study the thermo-particle transport  in the normal state of a Fermi gas with a tunable interaction. We calculate several thermo-transport quantities, including particle conductivity $G$,  thermal conductivity $G_T$, Seeback coefficient $\alpha$ and the dimensionless figure of merit $ZT=\frac{G\alpha^2 T}{G_T}$ , which is known to be the most important quantity  in determining the energy conversion efficiency between heat and particle currents.
We implement  the Nozi\`{e}res and Schmitt-Rink (NSR) scheme  to  consider the effect of preformed Cooper pairs\cite{NSR,Ohashi} and calculate  thermal transport quantities by Keldysh formulism \cite{Keldysh,Kamenev}. We find that the thermo-transport in the unitary region is characterized by several distinguished features which are absent or behave completely opposite to those  in the weak interacting regime. First,   the figure of merit has a giant enhancement  near the superfluid transition temperature in the unitary Fermi gas. Second, increasing the transmission coefficient of the point contact can increase the figure of merit. This behavior is exactly opposite to the one in the weak interaction Fermi gas, in which the figure of merit decreases as the transport ability of QPC increases \cite{Brantut}. Finally, the Lorentz number $L=\frac{G_T}{TG}$ of unitary Fermi gas violates the Wiedemann-Franz law \cite{Wiedemann,Kittel}, which signals  the breakdown of  the Fermi liquid. The deviation from the Wiedemann-Franz law also  increases as the transmission coefficient increases.  All these thermo-transport properties of the unitary Fermi gas can be attributed to  the existence of preformed Cooper pairs above the superfluid transition temperature. Our calculation  not only explains the observed experimental results, but also can be extended to determine the possible existence of preformed Cooper pairs in superconducting materials.

\begin{figure}[t]
\includegraphics[width=0.48\textwidth]{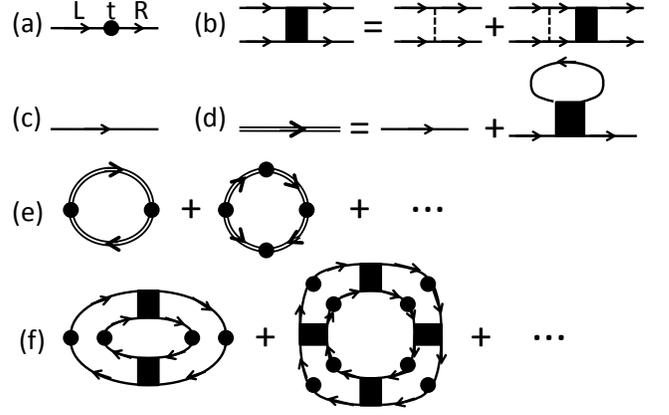}
\caption{The Feynman diagrams of (a), the tunneling term; (b), renormalized interaction under the NSR scheme; (c), free propagator of fermion; (d), full propagator of fermion under the NSR scheme; (e),current contributed by fermions; (f), current contributed by preformed Cooper pairs.}
\label{fig:feynman}
\end{figure}

\textit{Model and method.}--- We study the system with a two-terminal setup. Two reservoirs are connected by a quantum point contact, which is formed using high-resolution lithography \cite{Krinner1,Husmann,Krinner2}. A laser beam working as an attractive gate potential is shone on the QPC region. The number of the open channels can be changed by tuning the gate potential \cite{Nazarov}. In this work, we limit our calculation for the single open channel case. Particle and heat currents  are  generated by imposing temperature and chemical potential differences between the two reservoirs. The Hamiltonian of the system can be cast in three parts as the following(setting $\hbar=1$ and $k_B=1$)\cite{Cuevas,Bolech2004,Bolech2005},
\bea
\hat H=\hat H_{L}+\hat H_{R}+H_{T},
\eea
where $\hat H_L(\hat H_R)$ is the Hamiltonian of the left(right) reservoir and is written as
\bea
\hat H_{j}=&&\int d^3{\bf r}\Big\{\sum_{\sigma}\hat{\psi}^\dagger_{j\sigma}({\bf r})(-\frac{\nabla^2}{2m}-\mu_j)\hat\psi_{j\sigma}({\bf r})\cr &&-g\hat{\psi}^\dagger_{j\uparrow}({\bf r})\hat{\psi}^\dagger_{j\downarrow}({\bf r})\hat{\psi}_{j\downarrow}({\bf r})\hat{\psi}_{j\uparrow}({\bf r})\Big\}.
\eea
The operator $\hat\psi_{j\sigma}({\bf r})$ and $\hat\psi_{j\sigma}^\dagger({\bf r})$ describe the annihilation and creation of a fermion atom with spin $\sigma$ in the $j$-th reservoir with $j=L,R$. $m$ is the mass of fermions and $\mu_j$ is the chemical potential of the $j$-th reservoir. The parameter $g$ is the bare interaction strength between the spin up and spin down atoms. It's related to the $s$-wave scattering length $a_s$ by the renormalization relation $\frac{1}{g}=-\frac{m}{4\pi a_s}+\int \frac{d^3{\bf k}}{(2\pi)^3} \frac{1}{2\epsilon_{\bf k}}$, where $\epsilon_{\bf k}={\bf k}^2/2m$.   We study  this model  above the superfluid transition temperature under the NSR scheme, where the effect of preformed Cooper pairs is taken into account by calculating the ladder diagram as shown in Fig. \ref{fig:feynman} (b).

The Hamiltonian $\hat H_T$ describes the tunneling of fermions between the two reservoirs and is written as
\bea
\hat H_{T}=\sum_{\sigma}t \hat{\psi}^\dagger_{L\sigma}(0)\hat\psi_{R\sigma}(0)+h.c..
\eea Here we assume this process only occurs at a single point ${\bf x}=0$. In the single open channel case,  the transport properties through the QPC is assumed to be controlled  by the parameter $t$, the tunneling amplitude.

In the realistic cold atom experiment one reservoir may be heated up by a laser beam.  The heat and particle currents are generated between two reservoirs by temperature bias $\Delta T=T_L-T_R$ or chemical potential bias $\Delta \mu=\mu_L-\mu_R$.  They are defined as $I_Q\equiv\frac{1}{2}\langle\frac{\partial(\hat E_R-\hat E_L)}{\partial t}\rangle$ and $I_N\equiv\frac{1}{2}\langle\frac{\partial(\hat N_R-\hat N_L)}{\partial t}\rangle$, with $\hat E_{j}\equiv \sum_{\sigma}\int d^3{\bf r}\hat{\psi}^\dagger_{j\sigma}(t,{\bf r})i\partial_t\hat{\psi}_{j\sigma}(t,{\bf r})$ and $\hat N_{j}\equiv \sum_{\sigma}\int d^3{\bf r}\hat{\psi}^\dagger_{j\sigma}(t,{\bf r})\hat{\psi}_{j\sigma}(t,{\bf r})$ being the energy operator and number operator of the $j$-th reservoir respectively.
Using the Keldysh formulaism\cite{Keldysh,Kamenev}, the average $\langle...\rangle$ in above expressions is  calculated over the time-evolving many-body state on a closed time contour.
In the linear response regime,  we consider the small temperature and chemical potential bias, $\Delta T$ and $\Delta \mu$. The currents can be calculated up to the first order of $\Delta T$ and $\Delta \mu$ as the following,
\bea
\left(\begin{array}{cc} I_{N} \\ I_{Q} \end{array}\right)=\left(\begin{array}{cc}L_{11}& L_{12}\\ L_{21} & L_{22}\end{array} \right)\left(\begin{array}{cc} \Delta \mu \\ \Delta T \end{array}\right),\label{eq:current}
\eea
where $L_{ij}$ is known as the Lorentz matrix, the elements are given by
\bea
L_{11}= && -\int \frac{d\omega}{2\pi}\Big\{2\mathcal F^2(\omega)\frac{\partial n_F}{\partial \omega}  +4\mathcal B^2(\omega)\frac{\partial n_B}{\partial \omega}\Big\}, \cr
L_{12}=&& -\int \frac{d\omega }{2\pi}\frac{\omega}{\bar T}\Big\{2\mathcal F^2(\omega)\frac{\partial n_F}{\partial \omega}  +2\mathcal B^2(\omega)\frac{\partial n_B}{\partial \omega}\Big\}, \cr
L_{21}= && -\int \frac{d\omega }{2\pi}\omega\Big\{2\mathcal F^2(\omega)\frac{\partial n_F}{\partial \omega}  +2\mathcal B^2(\omega)\frac{\partial n_B}{\partial \omega}\Big\}, \cr
L_{22}= && -\int \frac{ d\omega}{2\pi} \frac{\omega^2}{\bar T}\Big\{2\mathcal F^2(\omega)\frac{\partial n_F}{\partial \omega}  +\mathcal B^2(\omega)\frac{\partial n_B}{\partial \omega}\Big\}.
\eea
The elements $L_{12}$ and $L_{21}$ are related by the Onsager relation as $L_{21}=\bar T L_{12}$ \cite{Onsager1,Onsager2}, where $\bar T=(T_L+T_R)/2$. $n_F(\omega)=\frac{1}{\exp(\beta\omega)+1}$ and $n_B(\omega)=\frac{1}{\exp(\beta\omega)-1}$ are the Fermi and Bose distribution functions, where $\beta=1/T$. In above expression one observes that both particle and heat currents include two parts. One is from fermions and the other one  is from bosons, which is the contribution of the preformed Cooper pairs. These two kinds of contributions can be summarized by the Feynman diagrams in Fig. \ref{fig:feynman} (e) and (f). Both of them are infinite summation to all the order of $t$. The function $\mathcal F(\omega)$ and $\mathcal B(\omega)$ are expressed as $\mathcal F(\omega)=\frac{2\mathcal T A_{F}(\omega)/A_{F}(0)}{\big|1+\mathcal T^2 (A_{F}(\omega))^2/(A_{F}(0))^2\big |}$ and $\mathcal B(\omega)=\frac{\pi \mathcal T^2\sqrt\beta A_B(\omega)}{\big|1- \mathcal T^4\beta ({ReG}_{B}(\omega))^2/4\big|}$, the Taylor expansion of which with respect to $\mathcal T$ manifestly demonstrates the infinite summation of graphs in  Fig. \ref{fig:feynman} (e) and (f). Here we define a dimensionless transmission amplitude $\mathcal T=\pi t A_F(0)$. With the recently developed technique of scanning gate microscope in cold atom experiments,  the effective tunneling amplitude $\mathcal T$ can be tuned continuously from 0 to 1 \cite{Hausler,Lebrat,Yao}. The spectral functions of fermions and Cooper pairs are given by
\bea
&&A_F(\omega)=-\frac{1}{\pi }\int \frac{d^2{\bf k}}{(2\pi)^2} {\rm Im}\big( \frac{1}{\omega+i 0^{+}-\epsilon_{\bf k}+\mu-\Sigma_F(\omega,{\bf k})}\big), \cr
&&A_B(\omega)=-\frac{1}{\pi} \int \frac{d^2 {\bf q}}{(2\pi)^2} {\rm Im}\big( G_B(\omega+i0^+,{\bf q})\big),\label{eq:spectral}
\eea
where $\Sigma_F(\omega,{\bf k})$ is the self energy of the Fermions $\Sigma_F(\omega,{\bf k})=-\frac{1}{\beta}\sum_{\omega_m}\int\frac{d^3{\bf p}}{(2\pi)^3}\frac{G_B(i\omega_m,{\bf p})}{[\omega+i0^+-i\omega_m+({\bf p}-{\bf k})^2/2m-\mu]}$, and the propagator of Cooper pair is $ G_B(i\omega_m,{\bf k})=\Big(-\frac{m}{4\pi a_s}-\int\frac{d^3{\bf p}}{(2\pi)^3}\Big[\frac{1-n_F(\epsilon_{\bf p}-\mu_j)-n_F(\epsilon_{{\bf p}-{\bf k}}-\mu)}{-i\omega_m+\epsilon_{\bf p}+\epsilon_{{\bf p}-{\bf k}}-2\mu}-\frac{1}{2\epsilon_{\bf p}}\Big]\Big)^{-1}.$
In the realistic cold atom experiments, the QPC structure is not directly connected to the three-dimensional reservoirs. Instead it's connected to a two-dimensional region first, which is formed by the repulsive potential of a TEM01-like mode of a laser \cite{Krinner1,Husmann,Krinner2}.  Hence, the integrations of momenta in the spectral function of Eq.(\ref{eq:spectral}) is two-dimensional. As illustrated in Ref. \cite{Liu2017} the low dimensional structure is very important for the enhancement of the conductance.

Usually, the currents in Eq. (\ref{eq:current}) are expressed in terms of the thermo-particle coefficients as the following,
\bea
\left(\begin{array}{cc} I_{N} \\ I_{Q} \end{array}\right)=G \left(\begin{array}{cc}1& \alpha \\ \bar T\alpha & \bar T(L+\alpha^2) \end{array} \right)\left(\begin{array}{cc} \Delta \mu \\ \Delta T \end{array}\right),
\eea
where the conductance $G$,  Seebeck coefficient $\alpha$ and the Lorentz number $L$ are defined as
$G=L_{11}$, $\alpha=L_{12}/L_{11}$, $L=L_{22}/L_{11}-(L_{12}/L_{11})^2$ and $\bar T=(T_L+T_R)/2$. Furthermore the thermal conductivity is defined as $G_T=\bar TGL$. Then the figure of merit can be calculated as
\bea
ZT=\frac{G \alpha^2 \bar T }{G_{T}}=\frac{L_{12}L_{21}}{L_{22}L_{11}-L_{12}L_{21}}.
\eea

\begin{figure}[t]
\includegraphics[width=0.48\textwidth]{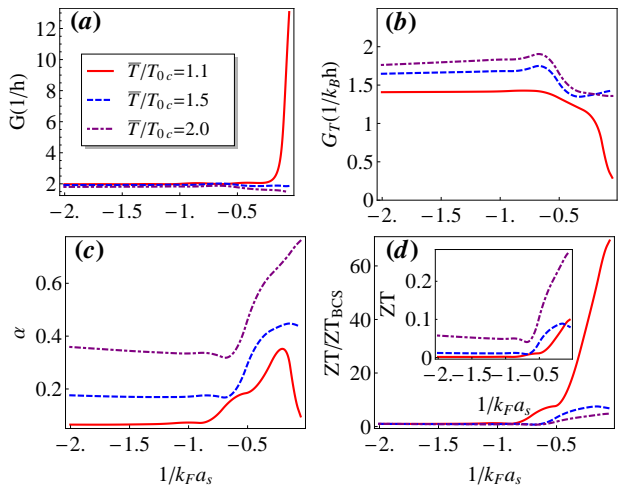}
\caption{(Color online) The thermo-particle coefficients (a), the conductance $G$; (b), the thermal conductance $G_T$; (c), the Seebeck coefficient $\alpha$, and (d), the ratio of figure of merit $ZT/ZT_{BCS}$ as functions of $1/a_sk_F$. The red, blue dashed and purple dashed dotted lines are for temperature $T/T_{0c}=1.1, 1.5, 2$, respectively, where $T_{0c}$ is the critical temperature at the unitarity. The transmission coefficient is set as $\mathcal T=1$.}
\label{fig:para_as}
\end{figure}
\textit{Enhanced figure of merit.}---To investigate how the preformed Cooper pairs affect the figure of merit,  we fix the temperature $\bar T$ slightly above the critical temperature $T_{0c}$ at $1/a_sk_F=0$, which is $T_{0c}=0.218T_F$ in the calculation of NSR, and study the variation of thermal coefficients with respect to the interaction strength.  The system becomes  closer to its critical temperature when $1/a_sk_F$  increases from the BCS limit to the unitarity. Consequently, the effect of preformed Cooper pairs is much stronger at the unitarity than  at the BCS limit. In Fig.\ref{fig:para_as}(a),  the conductance is  shown to be significantly enhanced when the unitarity is approached at a temperature very close to $T_{0c}$.  Due to the divergence of the Bose distribution function $1/(\exp(\beta\omega)-1)$ when the energy $\omega$ approaches $0$, the preformed Cooper pairs can generate a large contribution to the conductance between the two reservoirs \cite{Liu2017}. At the BCS limit, the conductance obeys the calculation of Landauer-B\"{u}ttiker formula~\cite{Landauer, Buttiker}, $G=2/h$ for one open channel case.

In contrast to the enhancement of particle conductance, the thermal conductance drops rapidly as one approaches the unitary regime as illustrated in Fig.\ref{fig:para_as} (b). The strong separation of particle and thermal transport timescales has been observed in the experiment of Ref. \cite{Husmann2018}, where they found the thermal conductance $G_T$ is roughly one order of magnitude smaller than the non-interacting case. Our calculation is  consistent with their results.

The Seebeck coefficient $\alpha$ demonstrates a non-monotonic behavior as a function of $1/a_sk_F$ as shown in Fig. \ref{fig:para_as} (c). The drop of $\alpha$ around the unitarity is due to the large enhancement of $G$ since $\alpha=L_{12}/G$. Although the Seebeck coefficient drops in the unitary regime, it can not overcome the combined effect of significantly increased conductance and decreased thermal conductance.  The combined effect results in an enhanced figure of merit in the unitary Fermi gas. In Fig. \ref{fig:para_as} (d) we compare ZT for different $1/a_sk_F$ with ZT$_{BCS}$, which is the figure of merit at $1/a_sk=-2$.  At the temperature $T/T_{0c}=1.1$, $ZT/ZT_{BCS}$  increases by about two order of magnitude at the unitarity case. As the temperature increases, the enhancement becomes weaker and weaker due to the decrease of the preformed Cooper pair fraction.

\begin{figure}[t]
\includegraphics[width=0.48\textwidth]{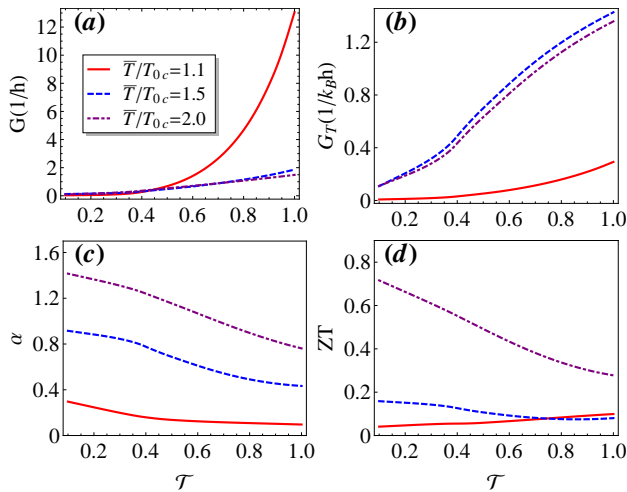}
\caption{(Color online) The thermo-particle coefficients (a), the conductance $G$; (b), the thermal conductance $G_T$; (c), the Seebeck coefficient $\alpha$, and (d), the figure of merit $ZT$ at $1/a_sk_F=0$ as functions of the transmission coefficient $\mathcal T$. The red, blue dashed and purple dashed dotted lines are for temperature $T/T_{0c}=1.1, 1.5, 2$, respectively.}
\label{fig:para_T}
\end{figure}
\textit{The variation of  the figure of merit by tuning the transmission amplitude $\mathcal T$}--- Recently a new technique in cold atom experiment has been developed, with which the transmission coefficient $\mathcal T$ can be continuously tuned from 0 to 1 by imprinting a mesoscopic potential or a lattice into the QPC tunneling channel\cite{Hausler,Lebrat,Yao}. Here we investigate the variation of the transport coefficients and  the figure of merit by tuning $\mathcal T$. The Fig.\ref{fig:para_T}(d) shows that $ZT$ increases as the $\mathcal T$ decreases for temperature $T/T_{0c}=1.5$ and $2$. In the experiment to study the thermo-particle effect of non-interacting Fermi gas\cite{Brantut}, it has been shown that the figure of merit increases as the disorder strength becomes stronger. Large disorder strength suppresses the transport ability.  Hence, the two results are consistent with each other. However, as the temperature decreases to $T/T_{0c}=1.1$ and the effect of preformed Cooper pairs becomes profound, the figure of merit $ZT$ presents a different behavior from the cases of $T/T_{0c}=1.5$ and $2$. Namely, $ZT$ increases as $\mathcal T$ approaches $1$. In Fig. \ref{fig:para_T} (b) and (c) one observes that $G_T$ increases and $\alpha$ decreases as $\mathcal T$  increases, which could result in the decrease of $ZT$. However, as shown in Fig. \ref{fig:para_T} (a) $G$ is significantly enhanced as $\mathcal T$ increases for $T/T_{0c}=1.1$. Eventually the behavior of $G$ becomes the dominant factor and results in the increase of $ZT$.

\begin{figure}[t]
\includegraphics[width=0.45\textwidth]{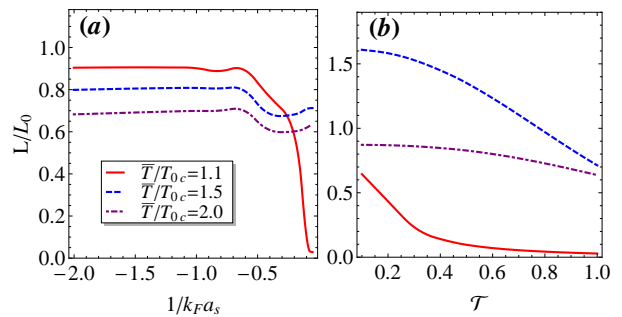}
\caption{(Color online) (a), the Lorentz number as functions of  $1/a_sk_F$ for $\mathcal T=1$;  (b), the Lorentz number as functions of  $\mathcal T$ for $1/a_sk_F=0$. The red, blue dashed and purple dashed dotted lines are for temperature $T/T_{0c}=1.1, 1.5, 2$, respectively.}
\label{fig:L}
\end{figure}

\textit{Violation to the Wiedemann-Franz law.}---In Fermi liquid,  the Lorentz number $L$ takes a universal value $L_0=\pi^2/3$ at low temperature. This is known as Wiedemann-Franz law\cite{Wiedemann,Kittel}. The violation to the Wiedemann-Franz law indicates a non-Fermi liquid behavior. In Fig.\ref{fig:L}(a), the Lorentz number is shown to be close to the universal value $L_0$ at the BCS limit at low temperature $T/T_{0c}=1.1$. However, it drops rapidly in the unitary regime, which is a clear signal of the breakdown of  Fermi liquid at the unitary regime.

Next, we investigate the variation of the Lorentz number at $1/a_sk_F=0$ with respect to the transmission coefficient $\mathcal T$. In Fig.\ref{fig:L}(b) it is demonstrated that $L$ increases and approaches the universal value $L_0$ as $\mathcal T$ decreases. We can conclude that the effect of preformed Cooper pairs becomes weaker as  $\mathcal T$ decreases. Namely, for a small transmission  coefficient,  the transport properties of unitary Fermi gases are closer to the Fermi liquid compared with the case of a large transmission  coefficient.

\textit{Concluding Remark.}---We theoretically investigate the thermo-particle transport properties of an unitary Fermi gas.  We find that the particle transport figure of merit can be enhanced dramatically in the unitary regime, comparing to the weak interaction regime.  The Wiedemann-Franz law is strongly violated as well. The underlying physics can be attributed to the existence of preformed Cooper pairs. These results not only explain recent experimental results but also suggest a new way to determine possible preformed Cooper pairs in other systems.

\textit{Acknowledgements}  We thank Hui Zhai and Shizhong Zhang for very helpful discussions. The work is supported by the Ministry of Science and Technology of China 973 program (No. 2015CB921300), National Science Foundation of China (Grant No. NSFC-1190020, 11534014, 11334012), and the Strategic Priority Research Program of CAS (Grant No.XDB07000000).

\end{document}